\documentclass[prb,aps,twocolumn]{revtex4}
\usepackage{epsfig,latexsym,amssymb,amsmath,amsbsy,graphics,graphicx,mathrsfs}

\begin{document}
\def \cation{$\kappa$-(ET)$_2$}~
\def \kpx{$\kappa$-(ET)$_2X$}
\def \kbr{$\kappa$-(ET)$_2$Cu[N(CN)$_2$]Br}
\def \deut8br{$\kappa$(d8)-(ET)$_2$Cu[N(CN)$_2$]Br}
\def \h8br{h8-(ET)$_2$Cu[N(CN)$_2$]Br}
\def \kcl{$\kappa$-(ET)$_2$Cu[N(CN)$_2$]Cl}
\def \kncs{$\kappa$-(ET)$_2$Cu(NCS)$_2$}
\def \kcn3{$\kappa$-(ET)$_2$Cu$_2$(CN)$_3$}
\def \>{\textgreater}
\def \<{\textless}
\def \q{\vec{q}}
\def \Q{\vec{Q}}
\def \kpcl{$\kappa$-Cl}
\def \kpbr{$\kappa$-Br}
\def \kpncs{$\kappa$-NCS}
\def \kpcn3{$\kappa$-(CN)$_3$}
\def \d8pbr{$\kappa$(d8)-Br}
\def \m{\mathrm{m}}
\def \max{\mathrm{max}}
\def \cross{\mathrm{cross}}
\def \M{\mathrm{M}}
\def \c{\mathrm{c}}
\def \lw{\mathrm{LW}}
\def \af{\mathrm{AF}}
\def \fm{\mathrm{FM}}
\def \sf{\mathrm{SF}}
\def \res{{\rho \sim T^2}}
\def \us{{\Delta v/v}}
\def \nmr{\mathrm{NMR}}
\def \ks{{K_s}}

\title{Antiferromagnetic Spin Fluctuations in the Metallic Phase
of Quasi-Two-Dimensional Organic Superconductors}
\author{Eddy Yusuf, B. J. Powell, and Ross H. McKenzie}
\affiliation{Department of Physics, University of Queensland, Brisbane,
Queensland 4072, Australia}
\date{\today}
\begin{abstract}

We give a quantitative analysis of the previously published nuclear
magnetic resonance (NMR) experiments in the \cation X family of organic
charge transfer salts. The temperature dependence of the nuclear spin
relaxation rate $1/T_1$, the Knight shift $K_s$, and the Korringa ratio
${\cal K}$ is compared to the predictions of the phenomenological spin
fluctuation model of Moriya, and Millis, Monien and Pines (M-MMP), that
has been used extensively to quantify antiferromagnetic spin
fluctuations in the cuprates. For temperatures above $T_\nmr \simeq 50$
K, the model gives a good quantitative description of the data in the
metallic phases of several \cation X materials. These materials display
antiferromagnetic correlation lengths which increase with decreasing
temperature and grow to several lattice constants by $T_\nmr$. It is
shown that the fact that the dimensionless Korringa ratio is much
larger than unity is inconsistent with a broad class of theoretical
models (such as dynamical mean-field theory) which neglects spatial
correlations and/or vertex corrections. For materials close to the Mott
insulating phase the nuclear spin relaxation rate, the Knight shift and
the Korringa ratio all decrease significantly with decreasing
temperature below $T_\nmr$. This cannot be described by the M-MMP model
and the most natural explanation is that a pseudogap, similar to that
observed in the underdoped cuprate superconductors, opens up in the
density of states below $T_\nmr$. Such a pseudogap has recently been
predicted to occur in the dimerised organic charge transfer salts
materials by the resonating valence bond (RVB) theory. We propose
specific new experiments on organic superconductors to elucidate these
issues. For example, measurements to see if high magnetic fields or
high pressures can be used to close the pseudogap would be extremely
valuable.

\end{abstract}

\maketitle

\section{Introduction}

 In the past twenty years a diverse range of new
 strongly correlated electron materials with exotic electronic and
magnetic properties have been synthesized. Examples include
 high-temperature cuprate superconductors,\cite{lee}
manganites with colossal magnetoresistance,\cite{dagotto:cmr}
 cerium oxide catalysts,\cite{esch}
sodium cobaltates,\cite{takada} ruthenates,\cite{
Meano&Mackenzie,grigera} heavy fermion materials,\cite{stewart} and
superconducting organic charge transfer salts.\cite{powell:review} Many
of these materials exhibit a subtle competition between diverse phases:
paramagnetic, superconducting, insulating, and the different types of
order associated with charge, spin, orbital, and lattice degrees of
freedom. These different phases can be explored by varying experimental
control parameters such as temperature, pressure, magnetic field, and
chemical composition. Although chemically and structurally diverse the
properties of these materials are determined by some common features;
such as, strong interactions between the electrons, reduced
dimensionality associated with a layered crystal structure, large
quantum fluctuations, and competing interactions. Many of these
materials are characterized by large antiferromagnetic spin
fluctuations.
 Nuclear magnetic resonance spectroscopy has proven to be a
 powerful probe of local spin dynamics in many
strongly correlated electron
materials.\cite{pennington,nmrnature,sidorov,miyagawa:chemrev} The
focus of this paper is on understanding what information about spin
fluctuations can be extracted from NMR experiments on the organic
charge transfer salts.

The systems which are the subject of the current study are the organic
charge transfer salts based on electron donor molecules BEDT-TTF (ET),
in particular the family \cation X (where $\kappa$ indicates a
particular polymorph\cite{ishiguro}). Similar physics occurs in the
other dimerised polymorphs, such as the $\beta$, $\beta'$, and
$\lambda$ phases.\cite{powell:review} These materials display a wide
variety of unconventional behaviours\cite{powell:review} including:
antiferromagnetic and spin liquid insulating states, unconventional
superconductivity, and the metallic phase which we focus on in this
paper. They also share highly anisotropic crystal and band structures.
However, for various sociological and historical reasons, the $\kappa$
salts have been far more extensively studied, and because we intend, in
this paper, to make detailed comparisons with experimental data, we
limit our study to $\kappa$ phase salts. This begs the question: do
similar phenomena to those described below occur in the $\beta$,
$\beta'$, or $\lambda$ salts? We would suggest that the answer is
probably \emph{yes} but this remains an inviting experimental question.

The metallic phase of \kpx~is very different from a conventional
metallic phase. Many features of the metallic phase agree well with the
predictions of dynamical mean field theory (DMFT)\cite{georges} which
describes the crossover from a `bad metal' at high temperatures to a
Fermi liquid as the temperature is
lowered.\cite{merino,hassan,limelette} This crossover from incoherent
to coherent intralayer\footnote{Throughout this paper when we discuss
coherent versus incoherent behavior we are discussing the behavior in
the planes unless otherwise stated. The subject of the coherence of
transport perpendicular to the layers is a fascinating issue. We refer
the interested reader to one of the reviews on the subject such as
Refs. \onlinecite{kartsovnik} and \onlinecite{Singleton-intra}.}
transport has been observed in a number of experiments such as
resistivity,\cite{limelette} thermopower,\cite{yu,merino} and
ultrasonic attenuation.\cite{frikach,fournier} The existence of
coherent quasiparticles is also apparent from the observed magnetic
quantum oscillations at low temperatures in
\kpx.\cite{singleton,wosnitza,kartsovnik} However, nuclear magnetic
resonance experiments (see Figs. \ref{fig:t1t_fit2} and
\ref{fig:korringa}) on the metallic phase on \kpx~are not consistent
with a Fermi liquid description. The nuclear spin relaxation rate per
unit temperature, $1/T_1T$, is larger than the Korringa form predicted
from Fermi liquid theory. As the temperature is lowered $1/T_1T$
reaches a maximum; we label this temperature $T_\nmr$ (the exact value
of $T_\nmr$ varies with the anion $X$, but typically,~$T_\nmr \sim 50$
K, see Fig. \ref{fig:t1t_fit2}). $1/T_1T$ decreases rapidly as the
temperature is lowered below $T_\nmr$ [see Fig
\ref{fig:t1t_fit2}].\cite{mayaffre,desoto,miyagawa:chemrev} The Knight
shift also drops rapidly around $T_\nmr$.\cite{desoto} This is clearly
in contrast to the Korringa-like behavior one would expect for a Fermi
liquid in which $1/T_1T$ and $K_s$ are constant for $T \ll T_F$, the
Fermi temperature. A similar non-Fermi liquid temperature dependence of
$1/T_1T$ and $K_s$ is observed in the cuprates.\cite{timusk,norman:adv}
It has been argued that the large enhancement of the measured $1/T_1T$
in cuprates is associated with the growth of antiferromagnetic spin
fluctuation within the CuO$_2$ planes as the temperature is
lowered.\cite{moriya,millis} The large decrease observed in $1/T_1T$
and $K_s$ measurements for underdoped cuprates\cite{timusk} at
temperatures well above $T_\c$ is suggestive of a depletion of the
density of states (DOS) at the Fermi level which might be expected if a
pseudogap opens at $T_\nmr$.

A quantitative description of spin fluctuations in the metallic phase
of \kpx~has not been given previously. However, the importance of spin
fluctuations for the superconducting \kpx~has been pointed out by
several
groups.\cite{powell:review,schmalian,kino,jujo,ben:prl,kuroki,powell:prl2007}
Since superconductivity arises from an instability of the metallic
phase, it is important to understand the strength of the spin
fluctuations in the metallic phase.

We use the phenomenological antiferromagnetic spin fluctuation model
which was first introduced by Moriya in his self consistent
renormalization (SCR) theory\cite{moriya} and then applied by Millis,
Monien and Pines (MMP)\cite{millis} to cuprates, to examine the role of
spin fluctuations in the metallic phase of \kpx. We fit the spin
fluctuation model to the nuclear spin relaxation rate per unit
temperature $1/T_1T$, Knight shift $K_s$, and Korringa ratio
$\mathcal{K}$ data. We find that the large enhancements measured in
$1/T_1T$ and $\mathcal{K}$ above $T_\nmr$ are the result of large
antiferromagnetic spin fluctuations [see Figs. \ref{fig:t1t_fit2} and
\ref{fig:korringa}]. The antiferromagnetic correlation length increases
as temperature decreases and the relevant correlation length is found
to be $2.8\pm1.8$ lattice spacings at $T=T_\nmr=50$ K. The model
produces a reasonable agreement with experimental data down to $T\sim
50$ K. The spin fluctuation model predicts a monotonically increasing
$1/T_1T$ with decreasing temperature while the measured $1/T_1T$ below
50 K is suppressed but never saturates to a constant value. This is
contrary to what is expected for a Fermi liquid where $1/T_1T$ is
constant. This indicates that the metallic phase of \kpx~is richer than
a renormalized Fermi liquid as has been previously thought to describe
the low temperature metallic state.

The structure of the paper is as follows. In Section II we introduce
the temperature dependence of the nuclear spin relaxation rate, Knight
shift, and Korringa ratio and describe how they probe the dynamic
susceptibility. We calculate these properties in a number of
approximations and contrast the results. In Section III we demonstrate
that the spin fluctuation model provides reasonable fits to the
existing experimental results for \kpx~above $T_\nmr$ and discuss its
limitations when applied to those materials. In Section IV we discuss
the unresolved issues and suggest new experiments to understand those
issues. Finally, we give our conclusions in Section V.

\section{The Spin Lattice Relaxation Rate,
Knight Shift, and Korringa Ratio}

In this section we discuss the temperature dependence of the nuclear
spin lattice relaxation rate $1/T_1$, Knight shift $K_s$, Korringa
ratio ${\cal K}$, and their dependence on the dynamic susceptibility
$\chi({\bf q},\omega) = \chi'({\bf q},\omega) + i \chi''({\bf
q},\omega)$. The general expressions for $1/T_1$, $K_s$ and ${\cal K}$
are given by\cite{barzykin}
\begin{subequations}
\label{nmr}
\begin{eqnarray}
\frac{1}{T_1} &=& \lim_{\omega \to 0}\frac{2k_BT}{\gamma_e^2\hbar^4}
\sum_{\bf q} |A({\bf q})|^2\frac{\chi''({\bf q},\omega)}{\omega},\label{t1t}\\
K_s &=& \frac{|A({\bf 0})| \chi'({\bf0},0)}{\gamma_e\gamma_N
\hbar^2}, \label{ks}\\ {\textrm{and}} \hspace{0.8cm} && \nonumber
\\\mathcal{K} &=& \frac{\hbar}{4\pi
k_B}\left(\frac{\gamma_e}{\gamma_N}\right)^2 \frac{1}{T_1T K_s^2},
\label{korringa}
\end{eqnarray}
\end{subequations}
where $A({\bf q})$ is the hyperfine coupling between the nuclear and
electron spins, and $\gamma_N$ ($\gamma_e$) is the nuclear (electronic)
gyromagnetic ratio. For simplicity we will consider a momentum
independent hyperfine coupling $|A|$ in what follows. Note that Eqs.
(\ref{nmr}) show that this is an approximation for $T_1$ but that it is
not an approximation at all for $K_s$. This is because $K_s$ only
probes the long wavelength physics and hence only depends on $A({\bf
0})$, the hyperfine coupling at ${\bf q}=\bf0$.

The calculation of the quantities in Eqs. (\ref{nmr}) boils down to
determining the appropriate form of the dynamic susceptibility. Below
we discuss, in some detail, the dynamic susceptibility within the spin
fluctuation model and calculate $1/T_1T$, $K_s$, and ${\cal K}$. The
results from dynamical mean field theory (DMFT) will also be discussed
for comparison.

\subsection{The Spin Fluctuation Model}\label{sect:sf}

The dynamic susceptibility in this model is given
by\cite{moriya,millis}
\begin{equation}
\chi({\bf q},\omega) = \chi_\lw(\omega) + \chi_\af({\bf q},\omega),
\label{dynamic}
\end{equation}
where $\chi_\lw(\omega)$ is the dynamic susceptibility in the long
wavelength regime and $\chi_\af({\bf q},\omega)$ is a contribution to
the dynamic susceptibility which is peaked at some wave vector ${\bf
Q}$. These susceptibilities take the form
\begin{eqnarray}
\chi_\lw(\omega) &=& \frac{\bar{\chi}_0(T)}{1-i\omega/\Gamma(T)}\nonumber\\
\chi_\af({\bf q},\omega) &=& \frac{\chi_Q(T)}{1+\xi(T)^2|{\bf q}-{\bf
Q}|^2-i\omega/\omega_\sf(T)}, \label{dynamic_af}
\end{eqnarray}
where $\bar{\chi}_0(T)$ [$\chi_Q(T)$] is the static spin susceptibility
at ${\bf q}={\bf 0}$ [${\bf Q}$], $\Gamma(T)$ [$\omega_\sf(T)$] is the
characteristic spin fluctuation energy which represents damping in the
system near ${\bf q}={\bf0}$ [${\bf Q}$], and $\xi(T)$ is the
temperature dependent correlation length. Hence, the real and imaginary
parts of the dynamic susceptibility can then be written as
\begin{eqnarray}
\chi'({\bf q},0) &=& \bar{\chi}_0(T) \left[1+\frac{\chi_Q(T)}
{\bar{\chi}_0(T)}\frac{1}{(1+\xi(T)^2|{\bf q}-{\bf Q}|^2)^2}\right]\nonumber\\
\chi''({\bf q},\omega) &=& \frac{\omega
\bar{\chi}_0(T)}{\Gamma}\nonumber\\
&& \left[1+\frac{\chi_Q(T) \Gamma}{\bar{\chi}_0(T) \omega_\sf(T)}
\frac{1}{(1+\xi(T)^2|{\bf q}-{\bf Q}|^2)^2}\right].
\nonumber\\\label{real_im_af}
\end{eqnarray}
Note that the above form of $\chi_{LW}(\omega)$ is the appropriate form
for a Fermi liquid. Therefore, if the system under discussion is not a
Fermi liquid then the validity of this expression for
$\chi_{LW}(\omega)$ cannot be guaranteed. For example, the marginal
Fermi liquid theory predicts a different frequency
dependence.\cite{varma} If the dynamic susceptibility has a large peak
at ${\bf Q} \ne {\bf 0}$ then $1/T_1$ will not be strongly dependent on
the long wavelength physics [because $1/T_1$ measures the
susceptibility over the entire Brillouin zone, c.f., Eq. (\ref{t1t}),
and therefore will be dominated by the physics at ${\bf q}=\bf Q$]. On
the other hand, the Knight shift is a measure of the long wavelength
properties [c.f., Eq. (\ref{ks})] and therefore may be sensitive to the
details of $\chi_\lw(\omega)$. Below we follow MMP\cite{millis} and
explicitly assume that the uniform susceptibility ($\bar{\chi_0}$) and
the spin fluctuation energy near ${\bf q=0}$ ($\Gamma$) are temperature
independent. One justification for this approximation in organics is
that the Knight shift is not strongly temperature
dependent.\cite{desoto} However, this approximation breaks down in
systems where the uniform susceptibility is strongly temperature
dependent such as YBa$_2$Cu$_3$O$_{6.63}$\cite{monien:prb43} and
La$_{1.8}$Sr$_{0.15}$CuO$_4$.\cite{monien:prb43b}

In the critical region $\xi(T) \gg a$, where $a$ is the lattice
constant, one has\cite{millis}
\begin{eqnarray}
\chi_Q(T) &=& \left(\frac{\xi(T)}{\xi_0}\right)^{2-\eta}
\bar{\chi_0}\nonumber\\
\omega_\sf(T) &=& \left(\frac{\xi_0}{\xi(T)}\right)^{z}\Gamma,
\end{eqnarray}
where $\eta$ is the critical exponent which governs the power-law decay
of the spin correlation function at the critical point, $z$ is the
dynamical critical exponent, and $\xi_0$ is a temperature independent
length scale. The simplest assumptions are relaxation dynamics for the
spin fluctuations (characterized by $z=2$) and mean field scaling of
the spin correlations ($\eta=0$). Within these approximations the real
and imaginary parts of the dynamic susceptibility are given by
\begin{eqnarray}
\chi'({\bf q},0)&=&\bar{\chi}_0 \left[1+\sqrt{\beta}\frac{[\xi(T)/a]^2}
{[1+\xi(T)^2|{\bf q}-{\bf Q}|^2]^2}\right]\nonumber\\
 \chi''({\bf q},\omega) &=& \frac{\omega \bar{\chi}_0}{\Gamma}
\left[1+\beta \frac{[\xi(T)/a]^4}{[1+\xi(T)^2|{\bf q}-{\bf
Q}|^2]^2}\right],
\end{eqnarray}
where $\beta=(a/\xi_0)^4$. The temperature independent, dimensionless
parameter $\beta$ can also be expressed in terms of the original
variables appearing in the dynamic susceptibility in Eq.
(\ref{dynamic_af}) as
\begin{equation}
\beta=\frac{\chi_Q(T) \Gamma}{\bar{\chi}_0 \omega_\sf(T)}
\left(\frac{a}{\xi(T)}\right)^4.
\end{equation}
Written in this form, $\beta$ has a clear interpretation: it represents
the strength of the spin fluctuations at the wave vector $\bf Q$
relative to those at ${\bf q=0}$. We will now consider two cases:
antiferromagnetic and ferromagnetic spin fluctuations.

\subsubsection{Antiferromagnetic Spin Fluctuations}

If we have antiferromagnetic spin fluctuations then the dynamic
susceptibility $\chi({\bf q},\omega)$ is peaked at a finite wave vector
${\bf q}={\bf Q}$; for example, on a square lattice with nearest
neighbor exchange only, ${\bf Q}=(\pi,\pi)$. The NMR relaxation rate,
Knight shift, and Korringa ratio can be calculated straightforwardly
from the real and imaginary parts of the dynamic susceptibility given
in Eq. (\ref{real_im_af}). The results are
\begin{subequations}
\label{nmr_af}
\begin{eqnarray}
\frac{1}{T_1T} &=&\frac{2\pi k_B |A|^2
\bar{\chi}_0}{\gamma_e^2\hbar^4\Gamma}
\left[1+\beta\frac{[\xi(T)/a]^4}
{1+[\tilde{Q}\xi(T)]^2}\right]\label{nmr_af_t1t}\\
K_s &=& \frac{|A|\bar{\chi}_0}{\gamma_e \gamma_N \hbar^2}\left[1 +
\sqrt{\beta} \frac{[\xi(T)/a]^2}{1+[\tilde{Q}\xi(T)]^2}\right]
\label{nmr_af_ks}\\
{\cal K} &=&
\frac{\hbar\gamma_e^2}{2\Gamma\bar{\chi}_0}\frac{\left[1+\beta\frac{[\xi(T)/a]^4}
{1+[\tilde{Q}\xi(T)]^2}\right]}{\left[1 + \sqrt{\beta}
\frac{[\xi(T)/a]^2}{1+[\tilde{Q}\xi(T)]^2}\right]^2},\label{nmr_af_korringa}
\end{eqnarray}
\end{subequations}
where $\tilde{Q}$ is a cutoff from the momentum integration [c.f. Eq.
(\ref{t1t})]. For $\xi(T) \gg a$: $1/T_1T \sim \xi(T)^2$, and $K_s
\sim$ constant which leads to the Korringa ratio $\mathcal{K} \simeq
(\hbar\gamma_e^2/2\Gamma\bar{\chi}_0) [\tilde{Q}\xi(T)]^2$. In this
model the Korringa ratio can only be equal to unity if the spin
fluctuations are completely suppressed ($\beta=0$). Hence, one expects
${\cal K}>1$ if antiferromagnetic fluctuations are
dominant.\cite{moriya:jpsj,narath} It has been shown\cite{ypm}, that
the Korringa ratio is unity when the hyperfine coupling $A({\bf q})$ is
momentum independent and the vertex corrections are negligible. The
fact that the Korringa ratio is larger than one indicates that there
are significant vertex corrections when there are large
antiferromagnetic fluctuations.

\subsubsection{Ferromagnetic Spin Fluctuations}\label{sect:ferro}

For ferromagnetic spin fluctuations, $\chi({\bf q},\omega)$ is peaked
at ${\bf q}=\bf0$. The NMR relaxation rate is exactly the same as that
given in Eq. (\ref{nmr_af_t1t}) because $1/T_1T$ comes from summing the
contributions form all wave vectors in the first Brillouin zone, which
makes the location of the peak in $\chi({\bf q}, \omega)$ in the
momentum space irrelevant. In contrast, the Knight shift will be
different in the ferromagnetic and antiferromagnetic cases because
$K_s$ only measures the ${\bf q}=\bf0$ part of the dynamic
susceptibility; hence $K_s$ will be enhanced by the ferromagnetic
fluctuations. Thus, for ferromagnetic spin fluctuation description the
Knight shift $K_s$ is given by
\begin{eqnarray}
K_s &=& \frac{|A|\bar{\chi}_0}{\gamma_e \gamma_N \hbar^2}\left[1 +
\sqrt{\beta} (\xi/a)^2\right]\label{nmr_fm}
\end{eqnarray}
and the corresponding Korringa ratio by
\begin{eqnarray}
{\cal K} &=& \frac{\hbar\gamma_e^2}{2\Gamma\bar{\chi}_0}
\frac{\left[1+\beta\frac{[\xi(T)/a]^4}{1+[\tilde{Q}\xi(T)]^2}\right]}{\left[1
+ \sqrt{\beta} (\xi/a)^2\right]^2}.\label{nmr_fm_korringa}
\end{eqnarray}
For $\xi(T) \gg a$: $1/T_1T \sim \xi(T)^2$, and $K_s \sim \xi(T)^2$
which leads to  $\mathcal{K} \simeq
(\hbar\gamma_e^2/2\Gamma\bar{\chi}_0) [\pi\xi(T)/a]^{-2}$. Thus we see
that $\mathcal{K}<1$ in the presence of ferromagnetic
fluctuations.\cite{moriya:jpsj,narath} So again vertex corrections are
important if the system has strong ferromagnetic fluctuations. Recall
that, in contrast, for antiferromagnetic fluctuations the Korringa
ratio is larger than one. Thus evaluating the Korringa ratio allows one
to determine whether antiferromagnetic or ferromagnetic spin
fluctuations are dominant.

\subsection{Dynamical Mean Field Theory}\label{sect:dmft}

DMFT is an approach based on a mapping of the Hubbard model onto a
self-consistently embedded Anderson impurity
model.\cite{georges,kotliar,pruschke} DMFT predicts that the metallic
phase of the Hubbard model has two regimes with a crossover from one to
the other at a temperature $T_0$. For $T\<T_0$ the system is a
renormalized Fermi liquid characterized by Korringa-like temperature
dependence of $1/T_1T$ and coherent intralayer transport. Above $T_0$,
the system exhibits anomalous properties with $1/T_1T \sim a +
b(T_0/T)$ (c.f., Ref. \onlinecite{pruschke}) and incoherent charge
transport. This regime is often refereed to as the `bad
metal'.\cite{powell:review,merino} Microscopically the bad metal is
characterized by quasi-localized electrons and the absence of
quasiparticles.
This temperature dependence is similar to that for the single impurity
Anderson model.\cite{jarrell} Note that this temperature dependence is
similar to that found for spin fluctuations [c.f., Eq.
(\ref{limiting_t1t})].

The predictions of DMFT correctly describe the properties of a range of
transport and thermodynamic experiments on the organic charge transfer
salts.\cite{merino,hassan,limelette,analytis,powell:review} This
suggests that these systems undergo a crossover from a bad metal regime
for $T \> T_0$ to a renormalized Fermi liquid below $T_0$. However, we
will show below (also see Fig. \ref{fig:t1t_fit2}) that the nuclear
spin relaxation rate is suppressed but never saturates below $T_\nmr$;
this is \emph{not} captured by DMFT. This suggests that the
low-temperature regime of \kpx~is more complicated than the
renormalized Fermi liquid predicted by DMFT which, until now, has been
widely believed to be the correct description of the low temperature
metallic state in the organic charge transfer salts.

\section{Spin Fluctuations in \kpx}

The NMR relaxation rate, Knight shift, and Korringa ratio in the
antiferromagnetic spin fluctuations model are given by Eqs.
(\ref{nmr_af_t1t}), (\ref{nmr_af_ks}), and (\ref{nmr_af_korringa}).
Their temperature dependence comes through the antiferromagnetic
correlation length. We adopt the form of $\xi(T)$ from
M-MMP\cite{moriya,millis}: $\xi(T)/\xi(T_x) = \sqrt{2T_x/(T+T_x)}$. For
this form of the correlation length, $T_x$ represents a characteristic
temperature scale of the spin fluctuations and $\xi(T)$ is only weakly
temperature dependent for $T\ll T_x$. For this choice of $\xi(T)$ we
have
\begin{subequations}
\begin{eqnarray}
\frac{(T_1T)_0}{T_1T} &=& \left[1+\frac{\beta
C^2}{(T/T_x+1)^2+2\pi^2C(T/T_x+1)}\right]\hspace{25pt}\label{nmr_t1t}\\
K_s &=& (K_s)_0\left[1+\frac{\sqrt{\beta} C
}{1+2\pi^2C+T/T_x}\right]\label{nmr_ks}\\
{\cal K} &=& {\cal K}_0 \frac{\left[1+\frac{\beta
C^2}{(T/T_x+1)^2+2\pi^2C(T/T_x+1)}\right]}{\left[1+\frac{\sqrt{\beta} C
}{1+2\pi^2C+T/T_x}\right]^2},\label{nmr_korringa}
 \label{nmr_af2}
\end{eqnarray}
\end{subequations}
where we have defined
\begin{eqnarray}
C&=&2\left[\frac{\xi(T_x)}{a}\right]^2,\nonumber\\
(1/T_1T)_0&=&\frac{2\pi k_B |A|^2 \bar{\chi}_0}{\gamma_e^2\hbar^4\Gamma},\nonumber\\
(K_s)_0&=&\frac{|A|\bar{\chi}_0}{\gamma_e \gamma_N
\hbar^2},\nonumber\\ \textrm{and}~~ {\cal K}_0 &=&
\frac{\hbar\gamma_e^2}{2\Gamma\bar{\chi}_0},\label{coefficient}
\end{eqnarray}
to simplify the notation.

\subsection{The Nuclear Spin Relaxation Rate}

We now analyze the temperature dependence of $1/T_1$. In the discussion
to follow, we will assume that the correlation length is sufficiently
large compared to the lattice spacing and that the quantity $2\pi^2 C =
4\pi^2 [\xi(T_x)/a]^2$ is much larger than $T/T_x$. These two
assumptions imply that $2\pi^2 C (T/T_x+1)$ is more dominant than
$(T/T_x+1)^2$ in the denominator of the second term inside the square
bracket of Eq. (\ref{nmr_t1t}). Keeping only the dominant term, we
arrive at the expression for $1/T_1T$
\begin{equation}
 \frac{(T_1T)_0}{T_1T} \simeq 1 + \frac{\beta}{\pi^2}
 \left(\frac{\xi(T_x)}{a}\right)^2
\left(\frac{1}{T/T_x+1}\right). \label{limiting_t1t}
\end{equation}
The assumption $2\pi^2 C (T/T_x+1) \gg (T/T_x+1)^2$ can be easily
worked out to give a self consistency condition
\begin{equation}
\left[\frac{\xi(T_x)}{a}\right]^2 \gg \frac{T/T_x+1}{4\pi^2}.
\label{selfconsistency}
\end{equation}
We will use this relation later in Section IIIB as one of the tests for
the validity of our approximation.

\begin{figure*}
\epsfig{file=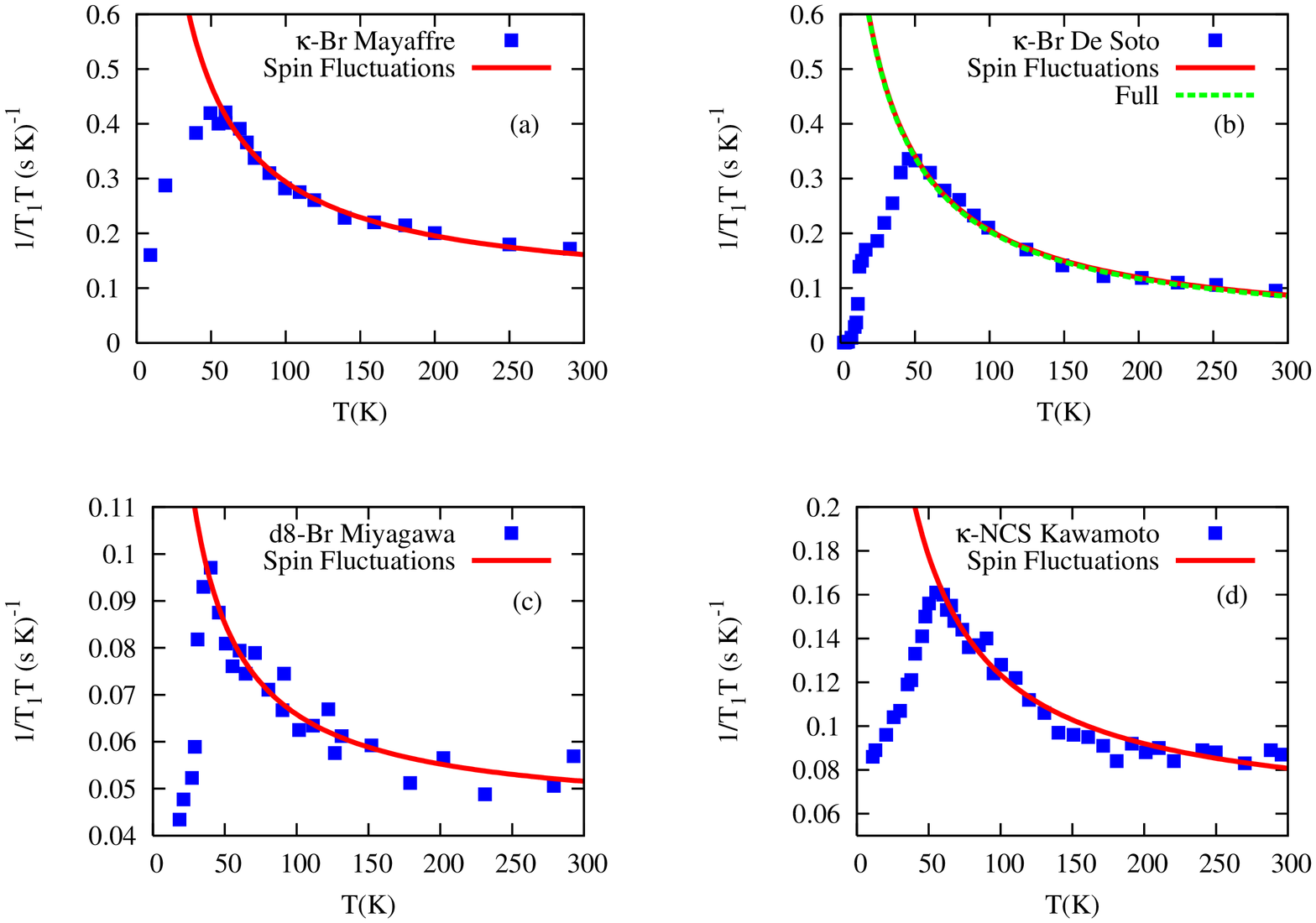,angle=-90,scale=0.6} \caption{[Color
online] Comparison of the measured nuclear spin relaxation rate per
unit temperature, $1/T_1T$, with the predictions of the spin
fluctuation model for various organic charge transfer salts. Panel (a)
shows data for \kbr~measured by Mayaffre {\it et al.}\cite{mayaffre}.
Panel (b) shows data for \kbr~measured by De Soto {\it et
al.}\cite{desoto}. Panel (c) shows data for \deut8br measured by
Miyagawa {\it et al.}\cite{miyagawa}. Panel (d) shows data for a
\kncs~powder sample measured by Kawamoto {\it et al.}\cite{kawamoto}
The $1/T_1T$ data are weakly temperature dependent at high
temperatures, have a maximum at $T_\nmr \sim 50$ K, and drop abruptly
below $T_\nmr\sim 50$ K, contrary to what one would expect for a Fermi
liquid in which $1/T_1T$ is constant. The remarkable similarities of
these data result from the quantitative and qualitative similarity of
the antiferromagnetic spin fluctuations in the metallic phases of these
materials. The parameters that produce the best fits (solid lines) to
Eq. (\ref{nmr_af2}) are tabulated in Table \ref{tab:parameter}. The
spin fluctuation model gives a good fit to the experimental data
between $T_\nmr\sim50$ and room temperature which suggests strong spin
fluctuations in the metallic states of \kbr, \deut8br, and \kncs.
However, below $T_\nmr$ the spin fluctuation model does not describe
the data well, indicating that some other physics dominates over the
spin fluctuations physics. In each figure the solid line is obtained
from the approximate form for $1/T_1T$ given by Eq.
(\ref{limiting_t1t}). To check that this approximation is reasonable,
we also plot $1/T_1T$ without any approximation, given by Eq.
(\ref{nmr_t1t}), as a dashed line in panel (b). The full and dashed
lines cannot be distinguished until well below $T_\nmr$ and so we
concluded that the approximation is excellent in the relevant regime.
Note that the analysis on $1/T_1T$ cannot differentiate between
antiferromagnetic and ferromagnetic spin fluctuations (see section
\ref{sect:ferro}), but the Korringa ratio strongly differentiates
between these two case and indicates that the fluctuations are
antiferromagnetic (see Fig. \ref{fig:korringa}).  The nomenclature
\kpbr, d8-Br, and \kpncs~is used as shorthand for \kbr, \deut8br, and
\kncs~respectively in the figure keys.} \label{fig:t1t_fit2}
\end{figure*}

The NMR relaxation rate per unit temperature calculated from the spin
fluctuation model [c.f., Eq. (\ref{nmr_af})] is a monotonic decreasing
function of temperature.
Thus one realizes immediately that the data, reproduced in Fig.
\ref{fig:t1t_fit2}, for temperatures below $T_\nmr$ is \emph{not}
consistent with the predictions of the spin fluctuation theory.
We will return to discuss this regime latter. We begin by investigating
the high temperature regime, $T>T_\nmr$.

We fit the $1/T_1T$ expression, Eq. (\ref{limiting_t1t}), to the
experimental data of De Soto\cite{desoto} for \kbr~ between $T_\nmr$
and 300 K with $(1/T_1T)_0$, $\beta [\xi(T_x)/a]^2$, and $T_x$ as free
parameters. It is not possible to obtain $\beta$ and $\xi(T_x)/a$
independently from fitting to $1/T_1T$ data because the model depends
sensitively only on the product $\beta [\xi(T_x)/a]^2$ (see Eq.
(\ref{limiting_t1t})). The results are plotted in Fig.
\ref{fig:t1t_fit2} and the parameters from the fits are tabulated in
Table \ref{tab:parameter}.
We have checked the validity of our approximation by plotting $1/T_1T$
given by Eq. (\ref{nmr_t1t}) for \kbr~in Fig. \ref{fig:t1t_fit2}b,
where there is Korringa ratio data (see Fig. \ref{fig:korringa}) and
thus we can determine $\beta$ and $\xi(T_x)/a$ individually. It can be
seen from Fig. \ref{fig:t1t_fit2}b that the disagreement between
$1/T_1T$ plotted from Eqs. (\ref{nmr_t1t}) and (\ref{limiting_t1t}) is
smaller than the thickness of the curves. Therefore, this approximation
is well justified. It will also be shown in Section IIIB that the
correlation length is indeed rather large and the self consistency
condition, Eq. (\ref{selfconsistency}), is satisfied, thus providing
further justification for the use of Eq. (\ref{limiting_t1t}) here.

The model produces a reasonably good fit to the experimental data on
\kbr~(Ref. \onlinecite{desoto}) between $T_\nmr$, the temperature at
which $1/T_1T$ is maximum, and room temperature. In the high
temperature regime (e.g., around room temperature), $1/T_1T$ has a very
weak temperature dependence, indicating weakly correlated spins. The
large enhancement of $1/T_1T$ can be understood in terms of the growth
of the spin fluctuations: as the system cools down, the spin-spin
correlations grow stronger which allows the nuclear spins to relax
faster by transferring energy to the rest of the spin degrees of
freedom via these spin fluctuations. Strong spin fluctuations, measured
by large values of $\beta [\xi(T_x)/a]^2$, are not only present in
\kbr~but also observed in other materials such as fully deuterated
\kbr~\{which will be denoted by \deut8br\} and \kncs. The results of
the fits for \deut8br~and \kncs~are shown in Fig. \ref{fig:t1t_fit2}.
The parameters that produce the best fits are also tabulated in Table
\ref{tab:parameter}. In all of the cases studied here, strong spin
fluctuations are evident from the large value of $\beta
[\xi(T_x)/a]^2$.

The nature of the spin fluctuations, i.e., whether they are
antiferromagnetic or ferromagnetic, cannot, even in principle, be
determined from the analysis on $1/T_1T$. Both cases yield the same
$1/T_1T$ [see Eq. (\ref{nmr_af_t1t}) and Sec II.B.2] because the
nuclear spin relaxation rate is obtained by summing all wave vector
contribution in the first Brillouin zone. However, in the next section
we will use the Korringa ratio to show that the spin fluctuations are
antiferromagnetic.

\begin{table*}
\begin{tabular}{c | c | c | c | c | c | c}
\hline\hline Material & Ref. & \hspace{1pt} $(1/T_1T)_0$
($s^{-1}$K$^{-1}$) \hspace{1pt} & \hspace{6pt} $T_x$ (K)\hspace{6pt} &
\hspace{2pt} $T_\nmr$ (K)\hspace{2pt} & \hspace{2pt}
$\beta[\xi(T_x)/a]^2$ \hspace{2pt} &
\hspace{3pt} $\xi(T_\nmr)/a$ \hspace{3pt}\\
\hline
\kpbr & Mayaffre [\onlinecite{mayaffre}] & 0.09 $\pm$ 0.01 & 7 $\pm 6$ & 60 & 290 $\pm 250$ & $?$\\
\kpbr& De Soto [\onlinecite{desoto}] & 0.02 $\pm$ 0.01 & 20 $\pm 10$ & 50 & 680 $\pm 430$ & $2.8\pm1.8$\\
d8-Br &Miyagawa [\onlinecite{miyagawa}] & 0.04 $\pm 0.01$ & 6 $\pm 4$ & 40 & 85 $\pm 65$ &$?$\\
\kpncs& \hspace{1pt} Kawamoto [\onlinecite{kawamoto}] \hspace{1pt}&
0.06 $\pm 0.01$ & 11 $\pm 3$ & 55 & 110 $\pm 90$ & $?$\\\hline\hline
\end{tabular}
\caption{The parameters obtained from the fits which are used to
produce Fig. \ref{fig:t1t_fit2}. Evidence for strong spin fluctuations
come from the large value of $\beta [\xi(T_x)/a]^2$ which are present
for all the materials tabulated above. $T_\nmr$ is determined from the
peak of $1/T_1T$ [see Fig. \ref{fig:t1t_fit2}]. The correlation length
shown in the last column in the table was obtained by analyzing the
Korringa ratio data available for \kbr~[see Section IIIB]. The
correlation length for Mayaffre \kpbr,\cite{mayaffre} Miyagawa
d8-Br,\cite{miyagawa} and Kawamoto's \kpncs\cite{kawamoto}~can not be
determined from our analysis because there are not sufficient data.
This is shown with question marks in the correlation length column. In
the table \kpbr, d8-Br, and \kpncs~are used as shorthand for \kbr,
\deut8br, and \kncs~respectively.} \label{tab:parameter}
\end{table*}

Below $T_\nmr$, the calculated $1/T_1T$ continues to rise while the
experimental data show a decrease in the nuclear spin relaxation rate
per unit temperature. However, the data do not reach a constant
$1/T_1T$ as expected for a Fermi liquid. This indicates that the
physics below $T_\nmr$ is dominated by some other mechanism not
captured by the spin fluctuation theory, Fermi liquid theory, or DMFT.

One might argue that the discrepancy between the theory and experiments
below $T_\nmr$ stems from our assumption of a q-independent hyperfine
coupling in the $1/T_1T$ expression. However, in section
\ref{knight-shift} we will show that the Knight shift is also
inconsistent with the predictions of the spin fluctuation model below
$T_\nmr$. While including the appropriate q-dependent hyperfine
coupling might change the temperature dependence of $1/T_1T$, it
certainly cannot affect the temperature dependence of the Knight shift
because $K_s$ only depends on $A({\bf 0})$ [as can be seen from Eq.
(\ref{ks})].

\subsection{The Korringa Ratio}

\begin{figure}
\begin{center}
\epsfig{file=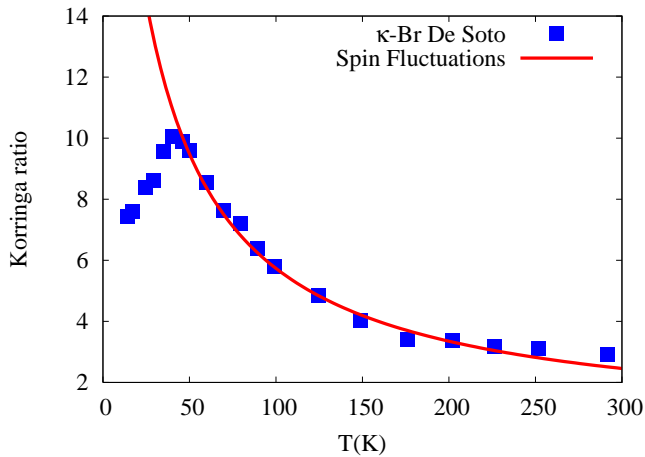,scale=0.7}
\end{center}
\caption{[Color online] Comparison of the Korringa ratio ${\cal
K}\propto1/T_1TK_s^2$ of \kbr~measured by De Soto {\it et
al}.\cite{desoto} with the prediction of the antiferromagnetic spin
fluctuation model. The best fit to Eq. (\ref{korringa-sf}) is indicated
by the solid line. The Korringa ratio is larger than 1 which indicates
that the spin fluctuations are antiferromagnetic (${\cal K}<1$ for
ferromagnetic fluctuations, see Section \ref{sect:ferro}). The
antiferromagnetic correlation length is found to be $2.8\pm1.8$ lattice
spacings at $T=50$ K. Below $T=50$ K the Korringa ratio is suppressed,
and the spin fluctuation model does not explain this behavior. This is
a clear indication that different physics is at play below 50 K.}
\label{fig:korringa}
\end{figure}

In the previous section we compared the predictions of the spin
fluctuation model for $1/T_1T$ to the experimental data and obtained
good agreement with the data between $T_\nmr$ and 300 K. However, we
were not able to determine $\beta$ and $\xi(T_x)/a$ independently
because $1/T_1T$ is sensitive only to the product $\beta
[\xi(T_x)/a]^2$. We were also unable to determine whether
antiferromagnetic or ferromagnetic spin fluctuations are dominant. We
resolve these questions by studying the Korringa ratio ${\cal K}$. It
has previously been pointed out that antiferromagnetic (ferromagnetic)
fluctuations produce a Korringa ratio that is larger (less) than
one.\cite{moriya:jpsj,narath} We have also seen in Section II that in
the limit of large correlation lengths, ${\cal K}\sim(\xi/a)^2 > 1$ for
antiferromagnetic spin fluctuations and ${\cal K}\sim(a/\xi)^2 < 1$ for
ferromagnetic spin fluctuations. The Korringa ratio data for \kbr~(see
Fig \ref{fig:korringa}) are significantly larger than one at all
temperatures which shows that antiferromagnetic fluctuations dominate.
With this in mind, we study the antiferromagnetic spin fluctuation
model.

First we note that $K_s$, given by Eq. (\ref{nmr_ks}), has a weak
temperature dependence because of our assumption that $[\xi(T_x)/a]$ is
generally larger than unity and $2\pi^2C \gg T/T_x$. Thus, the second
term inside the square bracket in Eq. (\ref{nmr_ks}) can be
approximated by $\sqrt{\beta}C/(1+2\pi^2C+T/T_x)^{-1} \simeq
\sqrt{\beta}/(2\pi^2)$ and the Knight shift will be given by $K_s
\simeq (K_s)_0 [1+\sqrt{\beta}/(2\pi^2)]$ which is temperature
independent. We use this temperature independent Knight shift to
calculate the Korringa ratio $\cal{K}$,
\begin{eqnarray}
{\cal K} &=& \frac{\hbar}{4\pi
k_B}\left(\frac{\gamma_e}{\gamma_N}\right)^2\frac{1}{T_1T K_s^2}
\label{korringa-sf}\\\nonumber &\simeq& {\cal
K}_0\left(1+\frac{\beta[\xi(T_x)/a]^2}
{\pi^2(T/T_x+1)}\right)\left(\frac{1}{1+\sqrt{\beta}/(2\pi^2)}\right)^2,
\end{eqnarray}
where the prefactor ${\cal K}_0$ is given by Eq.
(\ref{coefficient}).

We fit Eq. (\ref{korringa-sf}) to the experimental data for the
Korringa for \kbr.\cite{desoto} The result is plotted in Fig.
\ref{fig:korringa}. The Korringa ratio data are well reproduced by the
antiferromagnetic spin fluctuation model when $T>T_\nmr$. This is again
consistent with our earlier conclusion that the spin fluctuations are
antiferromagnetic. In this fit we have three free parameters, $\beta
[\xi(T_x)/a]^2$, $T_x$, and $\sqrt{\beta}$, two of which, $\beta
[\xi(T_x)/a]^2$ and $T_x$, have been determined from fitting $1/T_1T$.
There is only one remaining free parameter in the model,
$\sqrt{\beta}$, which can then be determined unambiguously from the
Korringa fit yielding $\beta=60 \pm 20$. This value of $\beta$ implies
that the antiferromagnetic correlation length $\xi(T) = 2.8 \pm 1.8 a$
($a$ is the unit of one lattice constant) at $T = 50$ K. This value is
in the same order of magnitude as the value of the correlation length
estimated in the cuprates.\cite{monien:prb43}

We now return to discuss the validity of our approximation which was
stated in the beginning of Section III A. The correlation length has
been determined to be $\xi(T)/a = 3.7\pm2.4$ at $T=T_x =20$ K from the
fit to the Korringa ratio data. This result surely satisfies the
requirement that the correlation length is larger than unit lattice
spacing. A stronger justification for our approximation comes from the
self consistency relation Eq. (\ref{selfconsistency}). With $\xi(T_x)/a
= 3.7$ and $T_x=20$ K, one could easily check that Eq.
(\ref{selfconsistency}) is indeed satisfied in the relevant regime,
i.e. between $T_\nmr$ and room temperature.

A large Korringa ratio\cite{takigawa:physica_c,bulut} has previously
been observed in the cuprates indicating similar antiferromagnetic
fluctuations in these systems. The Korringa ratio has also been
measured in a number of heavy fermion
compounds.\cite{ishida,aart,kitaoka} Similar antiferromagnetic
fluctuations are also present in CeCu$_2$Si$_2$; the Korringa ratio of
this material has a value of 4.6 at $T=100$ mK (Ref.
\onlinecite{aart}).
In contrast, YbRh$_2$Si$_2$\cite{ishida} and
CeRu$_2$Si$_2$\cite{kitaoka}, show strong ferromagnetic spin
fluctuations as is evident from the Korringa ratio less than unity. In
Sr$_2$RuO$_4$\cite{ishida:ruthenates} the Korringa ratio is
approximately 1.5 at $T=1.4$ K. Upon doping with Ca to form
Sr$_{2-x}$Ca$_x$$_2$RuO$_4$, the Korringa ratio becomes less than one
which indicates that there is a subtle competition between
antiferromagnetic and ferromagnetic fluctuations in these ruthenates.

\subsection{The Antiferromagnetic Correlation Length}

It is important to realize that the spin fluctuation formalism can be
used to extract quantitative information about the spin correlations
from NMR data.
From the fit for \kbr~(Table \ref{tab:parameter}) we found that the
antiferromagnetic correlation length $\xi(T)/a= 2.8 \pm 1.8$ at $T=50$
K. In order to understand the physical significance of this value of
$\xi(T)$ it is informative to compare this value with the correlation
length for the square\cite{ding} and triangular\cite{elstner} lattice
antiferromagnetic Heisenberg models with nearest neighbor interaction
only.

It has been shown\cite{ding} that, on the square lattice, the
antiferromagnetic Heisenberg model with nearest neighbor interaction
only has a correlation length of order $\xi(T)/a \sim 1$ for $T = J$
and of order $\xi(T)/a \sim 30$ for $T= 0.3 J$. On the other hand for
the antiferromagnetic Heisenberg model with nearest neighbor
interaction only on the isotropic triangular lattice, the correlation
length is only of order a lattice constant at $T= 0.3J$.\cite{elstner}
Thus the correlation length, $\xi(T)/a = 2.8\pm1.8$ at $T=50$ K,
obtained from the analysis of the data for \kbr~is reasonable and
places the materials between the square lattice and isotropic
triangular lattice antiferromagnetic Heisenberg model as has been
argued on the basis of electronic structure
calculations.\cite{mckenzie:comments,kino,powell:review}

One of the best ways to measure antiferromagnetic correlation length is
by inelastic neutron scattering experiments. To perform this
experiment, one needs high quality single crystals. Unfortunately, it
is difficult to grow sufficiently large single crystals for \kpx;
however, recently some significant progress has been made in this
direction.\cite{taniguchi-cry} Another way to probe the correlation
length is through the spin echo experiment. The spin echo decay rate
$1/T_2$ is proportional to the temperature dependence correlation
length. To the authors' knowledge there is no spin echo decay rate
measurement on the metallic phase of the layered organic materials at
the present time. Thus, it is very desirable to have such experimental
data to compare with the value of $\xi(T)$ we have extracted above.

\subsection{The Knight Shift}\label{knight-shift}

As we pointed out in Section II the Knight shift $K_s$ will generally
have a weak temperature dependence throughout the whole temperature
range and so, thus far, we have neglected its temperature dependence.
However, it is apparent from Eq. (\ref{nmr_af2}) that for any choice of
parameter values $\{\beta, \xi(T_x)/a$, and $T_x\}$, $K_s$ will always
increase monotically as the temperature decreases. Therefore the
temperature dependence of the Knight shift potentially provides an
important check on the validity of the spin fluctuation model. However,
in the following discussion one should recall the caveats (discussed in
section \ref{sect:sf}) on the validity of the calculation of the Knight
shift stemming from the assumption that the dynamics of the long
wavelength part of dynamical susceptibility relax in the same manner as
a Fermi liquid does.

In contrast to the prediction of the spin fluctuation model, the
experimental data \cite{desoto} for \kbr~show that $K_s$ decreases
slowly with decreasing temperature which then undergoes a large
suppression around $T_\ks \sim 50$ K. It should be emphasized here that
$T_\ks$ is approximately the same as $T_\nmr$, the temperature at which
$1/T_1T$ is maximum.

Since it is not possible to explain any of the NMR data below $T_\nmr$
in terms of the spin fluctuation model within the approximations
discussed thus far, we focus on the temperature range between 50 K to
300 K just as we did for the analysis of $1/T_1T$. Even in this
temperature range, there is a puzzling discrepancy between theory and
experiment: the experimental data decrease slowly with decreasing
temperature while the theoretical calculation predicts the opposite. We
will argue below that this discrepancy arises because the data are
obtained at constant pressure while the theoretical prediction assumes
constant volume. Since the organic charge transfer salts are
particularly soft, thermal expansion of the unit cell may produce a
sizeable effect to the Knight shift and may not be neglected. In
principle, an estimate of the size of this effect could be made
following Wzietek {\it et al.},\cite{wzietek} as
\begin{eqnarray}
\label{correction} \Delta K_s &=&
K_s^p-K_s^v\\&=&\nonumber\int_0^T{dT'}\left(\frac{\partial
K_s^p}{\partial P}\right)_{T'} \left(\frac{V\partial P}{\partial
V}\right)_{T'} \left(\frac{\partial V}{V\partial T'}\right)_P,
\end{eqnarray}
where $K_s^p$ is the (experimentally obtained) isobaric Knight shift,
$K_s^v$ is the (calculated) constant volume Knight shift, $(V
\partial P/\partial V)_T$ is the isothermal compressibility, and
$\frac{1}{V}(\partial V/ \partial T)_P$ is the linear thermal
expansion. However, it is not possible to obtain an accurate estimate
for $\Delta K_s$ at this time because there are no complete data sets
for $K_s^p$, isothermal compressibility, and thermal expansion as a
function of temperature and pressure for the \cation-X family. However,
a rough estimate for $\Delta K_s$ may be made using the available
experimental data.\cite{ypm} This suggests that the experimental data
is consistent with the spin fluctuation theory. Clearly, further
experiments are required to test this claim conclusively. Therefore we
raise this issue predominately to stress the importance of systematic
measurements of the parameters in Eq. (\ref{correction}).

Given the large uncertainty in $\Delta K_s$ we take $K_s$ to be
constant for temperatures above 50~K in the rest of this paper. This is
clearly the simplest assumption, it is not (yet) contradicted by
experimental data, and, perhaps most important, any temperature
dependence in the Knight shift is significantly smaller than the
temperature dependence of $1/T_1T$.

Regardless of the size of $\Delta K_s$, the Knight shift calculated
from the spin fluctuation model is inconsistent with the experimental
data below $T_\ks \sim 50$ K (see Fig. 4 in Ref. \onlinecite{ypm}). The
calculated $K_s$ shows a weakly increasing $K_s$ with decreasing
temperature, while the measured $K_s$ is heavily suppressed below 50 K.
One important point to emphasize here is that the temperature
dependence of $K_s$ will not change even if one uses the fully ${\bf
q}$-dependent $A({\bf q})$ since $K_s$ only probes the ${\bf q}=\bf0$
component of the hyperfine coupling and susceptibility [see Eq.
(\ref{ks})]. Thus, putting an appropriate q-dependent hyperfine
coupling will not change the result for $K_s$ (although it might give a
better description for $1/T_1T$). This provides a compelling clue that
some non-trivial mechanism is responsible to the suppression of
$1/T_1T$, $K_s$, and ${\cal K}$ below 50 K.

We have not addressed how the nuclear spin relaxation rate is modified
by the thermal expansion of the lattice. Since the organic compound is
soft, it is interesting to ask if there is a sizeable effect to
$1/T_1T$. Wzietek {\it et al.}\cite{wzietek} have performed this
analysis on quasi-1D organic compounds whose relaxation rate in found
to scale like $\chi_s^2$. One can straightforwardly derive the effect
of volume changes from the Hubbard model. If one uses the relation
$1/T_1T \sim \chi_s^2$ and assumes fixed $U$ and $t$, then $1/T_1T \sim
1/V^2$ will follow. However, it is clear from the phase diagram of the
organic charge transfer salts [see Ref. \onlinecite{powell:review}]
that there is a rather large change in $U$ and $t$ for even small
pressure variations. Therefore, there is no obvious relationship
between $1/T_1T$ and $\chi_s$ for the quasi-2D organics and it is not
clear how the imaginary part of the susceptibility $\chi''({\bf
q},\omega)$, which enters $1/T_1T$, is effected by thermal expansion
and lattice isothermal compressibility. Again, this stresses the
importance of the detailed experiments needed to determine the effect
of thermal expansion of the lattice on the measured relaxation rate.

\section{Open Problems and Future Experiments}

{\it Open problems.} The large suppression of $1/T_1T$ and $K_s$ below
$T_\nmr$ observed in all the $\kappa$ salts studied here cannot be
explained by the M-MMP spin fluctuation model. One plausible mechanism
to account for this feature is the appearance of a pseudogap which
causes the suppression of the density of states at the Fermi energy.
This is because at low temperature $1/T_1T$ and $K_s$ are proportional
to $\tilde{\rho}^2(E_F)$ and $\tilde{\rho}(E_F)$, where
$\tilde{\rho}(E_F$) is the full interacting density of states at the
Fermi energy.\cite{ypm} Independent evidence for the suppression of
density of states at the Fermi level comes from the linear coefficient
of the specific heat $\gamma$.\cite{timusk} The electronic specific
heat probes the density of excitations within $k_B T$ of the Fermi
energy. Any gap will suppress the density of states near the Fermi
surface which results in the depression of the specific heat
coefficient $\gamma$. Kanoda\cite{kanoda:jspj2006} compared $\gamma$
for several \kpx~salts and found that in the region close to the Mott
transition, $\gamma$ is indeed reduced. One possible interpretation of
this behavior is a pseudogap which becomes bigger as one approaches the
Mott transition. However, other interpretations are also possible. In
particular one needs to take care to account for the possible
coexistence of metallic and insulating phases; this is expected as the
Mott transition is first order in the organic charge transfer
salts.\cite{kagawa,sasaki} The existence of a pseudogap has also been
suggested in $\lambda$-(BEDT-TSF)$_2$GaCl$_4$\cite{suzuki} from
microwave conductivity measurements. The reduction of the real part of
the conductivity $\sigma_1$ from the Drude conductivity
$\sigma_\mathrm{dc}$ and the steep upturn in the imaginary part of the
conductivity $\sigma_2$ may be interpreted in terms of preformed pairs
leading to a pseudogap in this material. A pseudogap is predicted by
the RVB theory of organic
superconductivity.\cite{ben:prl,powell:prl2007}

The experimental evidence from measurements of $1/T_1T$, $K_s$, and
heat capacity all seem to point to the existence of a pseudogap below
$T_\nmr$ in \kbr~and \kncs. Thus a phenomenological description which
takes into account both the spin fluctuations which are important above
$T_\nmr$ and a pseudogap which dominates the physics below $T_\nmr$
would seem to be a reasonable starting point to explain the NMR data
for the entire temperature range (clearly superconductivity must also
be included for $T<T_\c$). We will pursue this approach in our future
work. In particular, if there is a pseudogap then important questions
to answer include: (i) How big is the pseudogap and what symmetry does
it have? (ii) How similar is the pseudogap in \kpx~to the pseudogaps in
the cuprates and in other strongly correlated materials such as
manganites and heavy fermions? (iii) Is there any relationship between
the pseudogap and the superconducting gap in \kpx? The answer to these
questions may help put constraints on the microscopic theories.

{\it Future experiments.} There are a number of key experiments
required to resolve the issue whether or not a pseudogap is present in
the low temperature metallic phase of \kpx. The pressure and magnetic
field dependence of the nuclear spin relaxation rate and Knight shift
will be valuable in determining the pseudogap phase boundary,
estimating the order of magnitude of the pseudogap, and addressing the
issue how the pseudogap is related to superconductivity. In the
cuprates, there have been several investigations of the magnetic field
dependence of the pseudogap seen in NMR experiments. For
Bi$_2$Sr$_{1.6}$La$_{0.4}$CuO$_6$ the nuclear spin relaxation rate does
not change with field up to 43 T.\cite{gqzheng2} However, since $T^*
\sim 200$ K, one may require a larger field to reduce the pseudogap.
Similar results were found in YBa$_2$Cu$_4$O$_8$.\cite{gqzheng}
However, in YBa$_2$Cu$_3$O$_{7-\delta}$ [see especially Fig. 6 of Ref.
\onlinecite{mitrovic}] a field of order 10 T is enough to start to
close the pseudogap.

The interlayer magnetoresistance of the cuprates has proven to be a
sensitive probe of the pseudogap.
\cite{mozorov,shibauchi,kawakami,elbaum} Moreover, it is found that for
the field parallel to the layers (which means that Zeeman effects will
dominate orbital magnetoresistance effects) the pseudogap is closed at
a field given by
\begin{equation}
H_{PG} \simeq \frac{k_B T^* }{\hbar  \gamma_e}
\end{equation}
where $T^*$ is the pseudogap temperature. For the hole doped cuprates
this field is of the order 100 T. In contrast, for the electron-doped
cuprates this field is of the order 30 T (and $T^* \sim 30-40$ K), and
so this is much more experimentally accessible.\cite{kawakami} The
field and temperature dependence of the interlayer resistance for
several superconducting organic charge transfer salts\cite{zuo} is
qualitatively similar to that for the cuprates. In particular, for
temperatures less than the zero-field transition temperature and fields
larger than the upper critical field, negative magnetoresistance is
observed for fields perpendicular to the layers. A possible explanation
is that, as in the cuprates, there is a suppression of the density of
states near the Fermi energy, and the associated pseudogap decreases
with increasing magnetic field.

A Nernst experiment can be used to probe whether there are
superconducting fluctuations in the pseudogap phase, as has been done
in the cuprates.\cite{wang} This experiment is particularly important
in understanding the relation between the pseudogap and
superconductivity.

One could also study the pressure dependence of the linear coefficient
of heat capacity $\gamma$. Since $\gamma$ is proportional to the
density of states at the Fermi energy, a detailed mapping of
$\gamma(P)$ would be an important probe for studying the pseudogap.
Finally, measurements of the Hall effect have also led to important
insights into the pseudogap of the cuprates\cite{timusk} and so perhaps
the time is ripe to revisit these experiments in the organic charge
transfer salts.

\section{Conclusions}

We have applied a spin fluctuation model to study the temperature
dependences of the nuclear spin relaxation rate, Knight shift, and
Korringa ratio in the metallic phase of several quasi two-dimensional
organic charge transfer salts. This model was based on Moriya's self
consistent renormalization theory\cite{moriya} which was then applied
by Millis, Monien, and Pines\cite{millis} to cuprates. The large
enhancement of $1/T_1T$ between $T_\nmr$ \{$\sim50$~K in \kbr\} and
room temperature has been shown to be the result of strong
antiferromagnetic spin fluctuations.

The antiferromagnetic correlation length is estimated to be $2.8\pm1.8$
lattice spacings in \kbr~at $T=50$ K. This value falls between those
for the Heisenberg model on the isotropic triangular lattice and the
square lattice.

The spin fluctuations in \kcl, \kbr, \deut8br, and \kncs~are found to
be similar both qualitatively and quantitatively. Strong spin
fluctuations seem to be manifested in materials close to Mott
transition. Recent NMR experiments\cite{kawamoto:ag} on \cation
Ag(CN)$_2\cdot$H$_2$O, which is situated further away from the Mott
transition, suggests that the spin fluctuations in this materials are
not as strong as those in the other $\kappa$ salts studied here.

The temperature dependence of $1/T_1T$ for $T \> T_\nmr$ from the spin
fluctuation model is qualitatively similar with the predictions of
dynamical mean field theory (DMFT). Below $T_\nmr$, the spin
fluctuation model predicts a monotically increasing $1/T_1T$ with
decreasing temperature while DMFT produces a constant $1/T_1T$. Neither
of these models can account for the large suppression of $1/T_1$,
$K_s$, and ${\cal K}$ below $T_\nmr \sim 50$ K observed in all the
$\kappa$-salts studied here. This suggests two things. First, the low
temperature regime is more complicated than the renormalized Fermi
liquid previously thought to be the correct description of the low
temperature metallic phase in these materials. Second, a pseudogap
exists at low temperatures near the Mott insulating phase of the
organic charge transfer salts.

\begin{acknowledgements}
The authors acknowledge stimulating discussions with Arzhang Ardavan,
Ujjual Divakar, John Fj\ae restad, David Graf, Anthony Jacko, Moon-Sun
Nam, David Pines, Rajiv Singh, and Pawel Wzietek. We are grateful to
Ujjual Divakar and David Graf for critically reading the manuscript.
This work was funded by the Australian Research Council.
\end{acknowledgements}


\end{document}